\documentclass[%
 twocolumn,
 amsmath,amssymb,
 pra,
longbibliography % for titles in references
]{revtex4-2}

\usepackage{tikz}
\usepackage{graphicx}% Include figure files
\usepackage{dcolumn}% Align table columns on decimal point
\usepackage{bm}% bold math
\usepackage{lipsum}
\usepackage{amsthm}
\usepackage{braket}
\usepackage[colorlinks=true,linkcolor=black,anchorcolor=black,citecolor=black,filecolor=black,menucolor=black,runcolor=black,urlcolor=black]{hyperref}
\usepackage[inline]{enumitem}
\usepackage[english]{babel}

\usepackage{algpseudocode}
\usepackage{algorithm}

\newtheorem{theorem}{Theorem}
\newtheorem*{theorem*}{Theorem}
\newtheorem{definition}[theorem]{Definition}
\newtheorem*{definition*}{Definition}

\newtheorem*{proposition*}{Proposition}
\newtheorem{corollary}[theorem]{Corollary}
\newtheorem*{corollary*}{Corollary}
\newtheorem{lemma}[theorem]{Lemma}
\newtheorem*{lemma*}{Lemma}

\newtheorem*{claim*}{Claim}

\newcommand{\norm}[1]{\left\lVert#1\right\rVert}
\newcommand{\R}{\mathbb{R}}
\newcommand{\C}{\mathbb{C}}
\newcommand{\I}{\mathbb{I}}
\DeclareMathOperator*{\argmax}{arg\,max}
\DeclareMathOperator*{\argmin}{arg\,min}

\newcommand{\ceil}[1]{\lceil #1 \rceil}

\definecolor{Arm}{rgb}{0.3,0.2,0.7}

\makeatletter
\def\blfootnote{\gdef\@thefnmark{}\@footnotetext}
\makeatother

\begin{document}

\preprint{APS/123-QED}

\title{Quantum Matching Pursuit: a Quantum Algorithm for Sparse Representations}

\author{Armando Bellante}
\affiliation{%
 Politecnico di Milano, DEIB, Via Ponzio 34/5 – Building 20, Milan 20133, Italy.
}%
\email{armando.bellante@polimi.it}

\author{Stefano Zanero}
\affiliation{%
 Politecnico di Milano, DEIB, Via Ponzio 34/5 – Building 20, Milan 20133, Italy.
}%

\date{\today}
\begin{abstract}
Representing signals with sparse vectors has a wide range of applications that range from image and video coding to shape representation and health monitoring.
In many applications with real-time requirements, or that deal with high-dimensional signals, the computational complexity of the encoder that finds the sparse representation plays an important role.
Quantum computing has recently shown promising speed-ups in many representation learning tasks. 
In this work, we propose a quantum version of the well-known matching pursuit algorithm. 
Assuming the availability of a fault-tolerant quantum random access memory, our quantum matching pursuit lowers the complexity of its classical counterpart of a polynomial factor, at the cost of some error in the computation of the inner products, enabling the computation of sparse representation of high-dimensional signals.
Besides proving the computational complexity of our new algorithm, we provide numerical experiments that show that its error is negligible in practice.
This work opens the path to further research on quantum algorithms for finding sparse representations, showing suitable quantum computing applications in signal processing.
\end{abstract}

\maketitle

%%%%%%%%%%%%%%%%%%%%%%%%%%%%%%%%%%%%%%%%%%%%%%%%%%%%%%%%%%%%%%%%%%%%%%%%%%%%%%%%%%%%%%%%%%%%%
%%%%%%%%%%%%%%%%%%%%%%%%%%%%%%%%%%%%%%%%%%%%%%%%%%%%%%%%%%%%%%%%%%%%%%%%%%%%%%%%%%%%%%%%%%%%%

\section{\label{sec:intro}Introduction}
Finding a sparse representation is the problem of representing a dense signal as a linear combination of a few unit vectors, also referred to as \emph{atoms}.
Usually, the set of atoms is larger than the space where the signal lies, as over-complete sets of atoms enable sparser representations~\citep{holger2008redundant}.
Once a set of atoms, or \emph{dictionary}, is fixed, the sparse representation of the signal is the set of coefficients of their linear combination.

Signals of the same type are likely to be represented sparsely over the same dictionary. 
For instance, the widely-used JPEG algorithm exploits the fact that images are sparse with respect to the Discrete Cosine Transform basis to perform compression~\citep{pennebaker1992jpeg}.
Finding sparse representations is a subject of interest in many fields, and its applications range from data compression to denoising and anomaly detection~\citep{elad2006image, adler2015sparse}.

When these applications have real-time requirements, or deal with high-dimensional signals, the computational cost of finding the representation is crucial. 
Unfortunately, finding the sparsest representation that approximates the signal is an NP-hard problem, and is intractable in practice.
For this reason, researchers have developed a series of greedy algorithms that, through local optimizations, find approximate solutions in an acceptable running time.

In recent years, the effectiveness of quantum computing in representation learning has become increasingly evident.
Recent research has proven computational advantages for algorithms such as principal component analysis~\citep{bellante2021quantum}, slow feature analysis~\citep{kerenidis2020classification}, and spectral clustering~\citep{kerenidis2021quantum}. 

In this work, we propose an end-to-end quantum algorithm for learning sparse representations using a matching pursuit approach.
We develop a quantum version of the well-known matching pursuit algorithm~\citep{mallat1993matching}, reducing its computational cost of a polynomial factor. While there are some known speed-ups in the case of specific analytical dictionaries~\citep{krstulovic2006mptk}, to our knowledge, there are no classical algorithms that compare with our run-time over a general dictionary.

Besides thoroughly analyzing the running time and error of our novel algorithm, we describe a suitable quantum processing framework that can be used as a starting point to construct other quantum pursuit algorithms.

The remainder of the paper is organized as follows. 
In Section \ref{sec:related}, we discuss previous work that relates to ours.
Section \ref{sec:classMP} describes the classical algorithm for matching pursuit, introducing the necessary notation for both the quantum and the classical versions. 
In Section \ref{sec:quantBackG}, we briefly introduce the concept of quantum computation and some subroutines that will serve as a basis for the novel quantum algorithm. 
Section \ref{sec:quantMP} presents the novel quantum matching pursuit algorithm, providing a thorough run-time and error analysis. 
Finally, in Section \ref{sec:experiments}, we run numerical experiments to show that the quantum matching pursuit can find representations that are as sparse as those of its classical counterpart.

%%%%%%%%%%%%%%%%%%%%%%%%%%%%%%%%%%%%%%%%%%%%%%%%%%%%%%%%%%%%%%%%%%%%%%%%%%%%%%%%%%%%%%%%%%%%%
%%%%%%%%%%%%%%%%%%%%%%%%%%%%%%%%%%%%%%%%%%%%%%%%%%%%%%%%%%%%%%%%%%%%%%%%%%%%%%%%%%%%%%%%%%%%%

\section{\label{sec:related}Related work}

The matching pursuit algorithm was first introduced by \citet{mallat1993matching}. The original version of the algorithm has a running time of $O(knm)$, where $k$ is the number of optimization iterations, $n$ is the length of the signal, and $m$ is the number of atoms in the dictionary.

Among the attempts to speed up the matching pursuit algorithm, the closest to ours is the one of \citet{krstulovic2006mptk}.
They exploited particular properties of some analytic (non-learned) dictionaries, like the multiscale time-frequency Gabor dictionary \citep{mallat1993matching}, to reduce the run-time of matching pursuit to $O(kn\log(n))$. 
However, their algorithm is still slow on non-analytical dictionaries.

While some previous works suggest the use of matching pursuit to simulate the dynamics of quantum mechanical processes \citep{wu2003matching, wu2004quantum, chen2006matching, wu2005matching}, to our knowledge there is no previous work that discusses quantum speed-ups for finding sparse representations of signals over large dictionaries.
%%%%%%%%%%%%%%%%%%%%%%%%%%%%%%%%%%%%%%%%%%%%%%%%%%%%%%%%%%%%%%%%%%%%%%%%%%%%%%%%%%%%%%%%%%%%%
%%%%%%%%%%%%%%%%%%%%%%%%%%%%%%%%%%%%%%%%%%%%%%%%%%%%%%%%%%%%%%%%%%%%%%%%%%%%%%%%%%%%%%%%%%%%%

\section{\label{sec:classMP}Classical Matching Pursuit}

\subsection{\label{subsec:notation}Notation}

We denote matrices using capital letters and use lowercase letters for vectors and scalars. 
Given a matrix $A$, its $i^{th}$ row and column are denoted by $a_{i,\cdot}$ and $a_i$ respectively.
The component identified by the $i^{th}$ row and the $j^{th}$ column is denoted $a_{ij}$. 
We write the $j^{th}$ element of a vector $u$ as $u_j$.

Let $x \in \C^n$ be a unit vector. 
Using Dirac's notation, we use $\ket{x}$ to represent it as a column vector and $\bra{x}$ to denote its complex-conjugate row vector.
We use $\braket{a_i,b_j}$ to denote the inner product between two vectors $a_i, b_j$.

The notation $\norm{\cdot}_2$ indicates the Euclidean norm of a vector. 
The pseudo-norm $\norm{\cdot}_0$ is the number of non-zero components of a vector. 
The symbol $\norm{\cdot}_F$ indicates the Frobenius norm of a matrix. 
For a matrix $A \in \R^{n \times m}$, the Frobenius norm is defined as $\norm{A}_F=\sqrt{\sum_i^n\sum_j^m a_{ij}^2}$.

When stating the complexity of an algorithm, the $\widetilde{O}()$ notation omits poly-logarithmic terms in the input data size (e.g., if an algorithm uses a matrix $A \in \R^{n \times m}$,  $\widetilde{O}(1) \equiv O(\text{polylog}(nm))$).

\subsection{Problem statement}
We can represent a signal as a vector $s \in \R^n$.
We use $d_j$ to denote the $j^{th}$ atom over which we search the sparse representation.
Each atom is a unit vector, meaning that for every $j$ we have $\norm{d_j}_2 = 1$.
A dictionary is a matrix $D \in \R^{n \times m}$ whose columns are the atoms $d_j$, for $j \in \{0, \dots, m-1\}$.
In most of the interesting cases, the dictionary is over-complete (i.e. $m>n$). 

Formally, given a signal $s \in \R^n$ and a dictionary $D \in \R^{n \times m}$, the problem of finding a sparse representation $x \in \R^m$ of the signal is known as $\mathcal{P}_0^\epsilon$.
\begin{definition}[Problem $\mathcal{P}_0^\epsilon$]
    Given $s \in \R^n, D \in \R^{n\times m}$, and $\epsilon \in \R^+$, problem $\mathcal{P}_0^\epsilon$ is defined as: 
    \begin{align}        
        \argmin_x \norm{x}_0 \text{such that} \norm{Dx-s}_2 \leq \epsilon
    \end{align}
\end{definition}

Finding the exact solution to problem $\mathcal{P}_0^\epsilon$ is an NP-hard task~\citep{natarajan1995sparse}. 
While quantum computers are not expected to solve NP-hard problems in polynomial time (indeed, it is widely believed that $\textsc{NP} \nsubseteq \textsc{BQP}$~\citep{bennett1997strengths}), they can still provide speed-ups of practical use on greedy algorithms that compute approximate solutions.
In this paper we propose a quantum version of the matching pursuit algorithm, a greedy approach to approximately solve the $\mathcal{P}_0^\epsilon$ problem in polynomial time.

\subsection{Algorithm}
The strategy behind the matching pursuit algorithm is to face the problem through subsequent optimization steps.
Starting from an empty solution $x = 0^{\otimes m}$, the matching pursuit searches for the atom that best reduces the difference between the representation $Dx$ and the signal at each iteration, updating the solution iteratively.
We now discuss the matching pursuit algorithm in detail.

As an initialization step, we create a residual vector and an empty solution
\begin{align}
    &r = s, \\
    &x = 0^{\otimes m}.
\end{align}
Since we are at the beginning of the algorithm and \\$s-Dx = s$, the residual is set equal to the signal.

Once the initialization is complete, the algorithm searches for the closest atom to the residual by computing
\begin{align}\label{eq:bestatom}
    j^* = \argmin_j \norm{r - z_jd_j}_2,
\end{align}
where $z_j \in \R$ is the best scaling factor for the atom $d_j              $
\begin{align}
    z_j = \argmin_z \norm{r - zd_j}.
\end{align}
It is possible to show that $z_j = \braket{r, d_j}$ \citep{mallat1993matching}, from which we can derive the following equivalence:
\begin{align}
    \label{eq:innercool}
    \norm{r - z_jd_j}_2^2 = \norm{r}_2^2 - |\braket{r, d_j}|^2
\end{align}
Because of Equation \ref{eq:innercool}, finding the best atom (Equation \ref{eq:bestatom}) is equivalent to searching for the maximum absolute value of the inner products between the current residual and the atoms
\begin{align}
    j^{*} = \argmax_j |\braket{r, d_j}|.
\end{align}
This step is known as the \emph{sweep stage}, as we need to iterate over all the atoms in the dictionary to compute the inner products and choose the best one.

After selecting the best atom, both the solution and the residual get updated
\begin{align}
    &x_{j^*} = x_{j^*} + z_{j^*},\\
    &r = r + z_{j^*}d_{j^*}.
\end{align}
Updating the residual makes it so that the algorithm does not consider the part of the signal that has been modeled so far.
At each iteration, the residual is $r = s - Dx$.

The algorithm continuously searches for the best approximating atom and performs the updates until the following stopping condition is met
\begin{align}
    \norm{x}_0 > L \text{ or } \norm{r}_2 \leq \epsilon, 
\end{align}
for a sparsity threshold $L \in \mathbb{N}^+$ and an error reconstruction tolerance $\epsilon \in \R^+$.
We remark that, at each iteration, the norm of the residual $\norm{r}_2 = \norm{s-Dx}_2$ expresses how well our solution approximates the original signal.

We summarize this procedure in Algorithm \ref{alg:mp}.
\begin{figure}[t]
\begin{algorithm}[H]
    \caption{Matching pursuit}
    \label{alg:mp}
    \begin{algorithmic}[1]
        \State \label{alg:MPinit}Initialize $r = s$, $x = 0^{\otimes m}$. 
        \While {not ($\norm{x}_0 > L$ or $\norm{r}_2 \leq \epsilon$)}
            \ForAll{$j \in [m]$} \label{alg:mpforstart}
                \State \label{alg:MPinner} Compute $\braket{d_j, r}$
            \EndFor\label{alg:mpforend}
            \State Select $j^* = argmax(\left|\braket{d_j, r}\right|)$
            \State Assign $z = \braket{d_{j^*}, r}$
            \State Update the solution $x_j = x_j + z$
            \State Update the residual $r = r - z d_{j^*}$
        \EndWhile
    \State Output $x$.
    \end{algorithmic}
\end{algorithm} 
\end{figure}

\subsection{Computational complexity}
The analysis of the run-time of the algorithm proceeds as follows.
The initialization step is linear in the length of the residual $O(n)$.
The computation of the sweep stage (steps \ref{alg:mpforstart}-\ref{alg:mpforend}) is the bottleneck of this algorithm. 
Indeed, the algorithm computes $m$ inner products of vectors of length $n$, which needs time $O(nm)$.
Selecting the best atom has a negligible cost, as it can be done during the computation of the inner products without significant overhead.
Updating the solution is $O(1)$ and the residual's update is bounded by $O(n)$.

The complexity of the sweep stage dominates all the other complexities in the loop.
Therefore, assuming that the matching pursuit converges after $k$ iteration, its asymptotic computational complexity scales as
\begin{align}
O(knm).
\end{align}

%%%%%%%%%%%%%%%%%%%%%%%%%%%%%%%%%%%%%%%%%%%%%%%%%%%%%%%%%%%%%%%%%%%%%%%%%%%%%%%%%%%%%%%%%%%%%
%%%%%%%%%%%%%%%%%%%%%%%%%%%%%%%%%%%%%%%%%%%%%%%%%%%%%%%%%%%%%%%%%%%%%%%%%%%%%%%%%%%%%%%%%%%%%

\section{\label{sec:quantBackG}Quantum Computing background}

\subsection{\label{subsec:quantinfo}Quantum computation}
Just like a bit is the fundamental unit of information in classical computing, a qubit is the fundamental information unit in quantum computing.
A qubit is a mathematical representation of a quantum mechanical object and can be described as an $\ell_2$ normalized vector of $\mathbb{C}^2$. 
The state of a $n$-qubit system (a register of a quantum computer) is the 
tensor product of single qubits: a unitary vector $\ket{x} \in H^{\otimes n} \simeq \mathbb{C}^{2^n}$.
In other words, with $\ket{i} \in H^{\otimes n}$ we denote a quantum register that contains the binary expansion of number $i$.
Its corresponding complex vector is a vector of length $2^n$, full of zeroes, with the $i^{th}$ element equal to one. 
For instance,
\begin{align}
    \ket{3} \in H^{\otimes 2} = \ket{1}\ket{1} = \begin{bmatrix}
        0\\
        0\\
        0\\
        1\\
    \end{bmatrix}.
\end{align}
Given a basis $\{\ket{i}\}_0^{n-1}$ for $H^{\otimes \log_2(n)}$, with $\log_2(n)$ qubits, we can describe a quantum state 
$\ket{\psi} = \sum_i^{n} \alpha_i \ket{i} $ with $\sum_i^{n} |\alpha_i|^2 = 1$.
The values $\alpha_i \in \mathbb{C}$ are called amplitudes of the quantum states $\ket{i}$ for the state $\ket{\psi}$. 

The evolution of a quantum system is described by unitary matrices $U$. 
Unitary matrices are norm-preserving and thus can be used as a suitable mathematical description of pure quantum evolutions. 
Any quantum algorithm that does not perform measurements can be represented by a unitary matrix.

Quantum states can be measured, but measurements alter the state. 
In this work, measurements are performed with respect to the computational basis $\{\ket{i}\}_0^{n-1}$ of $H^{\otimes \log_2(n)}$.
This means that if we measure a quantum register $\ket{\psi} = \sum_i^{n} \alpha_i \ket{i}$, it can collapse to any state $\ket{i}$, each with probability $|\alpha_i|^2$.
It is important to recall that no quantum algorithm can create a copy of a generic quantum state. 
Therefore, to measure a state multiple times, it is necessary to create it again from scratch every time.

For a deeper introduction to the subject, we encourage the reader to consult \citet{nielsenchuang}.

\subsection{\label{subsec:quantrandommem}Quantum random access memory}
A QRAM, or quantum ram,  is a device that: given a list of $n$ bit strings $x_i \in \{0,1\}^m$ of length $m$, performs the following mapping in time $O(\text{polylog}(n))$:
\begin{align}
    \sum_i^{n} \alpha_i \ket{i}\ket{0} \mapsto \sum_i^{n} \alpha_i\ket{i}\ket{x_i},
\end{align}
where $\alpha_i \geq 0$ and $\ket{x_i} \in H^{\otimes m}$ is the state of the computational basis that corresponds to the bit string $x_i$.

The reader can think of it as a quantum equivalent to the classical ram.
In a classical ram, we can store $n$ values and query any of those in time $O(1)$, considering it can access the $m$ bits in parallel.
The main difference with a classical ram is that a quantum ram needs to perform queries in superposition.

As explained in the next section, our quantum algorithm, like many previous ones,
assumes the availability of such a device to encode scalars, matrices, and vectors in quantum states efficiently.

Building a fault-tolerant, hardware-efficient QRAM is not an easy task. 
One of the most promising proposals to structure the quantum random access memory is the bucket-brigade architecture. 
First presented in \citet{giovannetti2008architectures}, this architecture is composed of $O(n)$ gates, while its circuit is only $O(\log(n))$ deep.
Current error analysis research claim that algorithms that query the bucket-brigade QRAM a super-polylogarithmic number of times (e.g., Grover's search, and consequently ours) likely require the bucket-brigade QRAM to be error corrected \cite{arunachalam2015robustness, hann2021resilience}.
This requires additional hardware and created skepticism on the effective speed-up of this class of algorithms in this input model \citep{arunachalam2015robustness}. 
Recently, \citet{hann2021resilience} have shown that the bucket-brigade architecture is highly resilient to generic errors and that its architectural advantage persists even in case of error correction, contrary to what was previously believed.
At the same time, a recent work has shown how to build a fault-tolerant bucket-brigade QRAM by parallelizing Clifford + T gates \cite{paler2020parallelizing}, at the cost of using $O(n)$ ancillary qubits.

In this work, we will perform our analysis assuming access to a fault-tolerant QRAM, capable of performing queries in time $O(\text{polylog}(n)) \sim \widetilde{O}(1)$, with $n$ the number of entries stored in the QRAM. 
With this assumption in mind, we can explain our data access model.

\subsection{\label{subsec:quantdatacces}Quantum data access}
We can encode a scalar $a \in \R$ in a quantum register $\ket{a}$ using its binary encoding and retrieve it by measuring the register in the computational basis, just as discussed in the previous section. 
To encode a scalar, we need as many qubits as the classical bits that store it.
For more detailed information on how to encode a real number in a quantum register, we suggest reading \citet[Chapter 5, Sections 5.1, 5.2]{nielsenchuang} to see how phase angles are encoded in a state, and \citet{rebentrost2021quantum} for a more formal definition of a quantum arithmetic model with fixed point precision.

On the other hand, the components of a vector $a \in \R^m$ can be encoded with fewer qubits than classical bits, using the amplitudes of a quantum state. 
When measured, the quantum state collapses to an index of the vector with probability proportional to the magnitude of the indexed component.
We call this representation state-vector. 

\begin{definition} [State-vector]
\label{Def:state_vector}
    Given a vector $x \in \R^n$, the corresponding state-vector is the following quantum state $\ket{x} = \frac{1}{\norm{x}_2}\sum_i^n x_i \ket{i}$, which is encoded in $\ceil{\log n}$ qubits.
\end{definition}

We say to have quantum access to a classical vector $x \in \R^n$ if we have access to a unitary operator that performs the mapping $U_x: \ket{0} \mapsto \ket{x}$ in time $O(log(n))$.
Similarly, we define the concept of quantum access to a matrix.

\begin{definition} [Quantum access to a matrix]
\label{Def:quantum_access}
    We say to have quantum access to a matrix $A \in \mathbb{R}^{n \times m}$ if we can perform the following mappings in time $O(\text{polylog}(nm))$:\\
    \begin{itemize*}
        \item $U: \ket{i} \ket{0} \mapsto \ket{i}\ket{a_{i, \cdot}} = \ket{i}\frac{1}{||a_{i,\cdot}||}\sum_j^m a_{ij}\ket{j}$, for $i \in \R^n$;
        \item $V: \ket{0} \mapsto \frac{1}{\|A\|_F}\sum_i^n \|a_{i,\cdot}\| \ket{i}$.
    \end{itemize*}
\end{definition}

By combining the two unitaries above, it is possible to represent a matrix in a quantum state 
\begin{align}\label{eq:matrice}
    \ket{A} = U(V\otimes \I)\ket{0}\ket{0} &= \frac{1}{\norm{A}_F}\sum_i^n\sum_j^m a_{ij}\ket{i}\ket{j} \\ 
    &= \frac{1}{\norm{A}_F}\sum_i^n \norm{a_{i,\cdot}}\ket{i}\ket{a_{i,\cdot}}
\end{align}
using two registers of $\ceil{\log(n)}+\ceil{\log(m)}$ qubits.

The appendix of \citet{kerenidis2017reccomender} shows in detail how to construct a classical data structure that enables efficient computation of the unitary matrices that give quantum access to vectors and matrices.
This classical data structure can be created in $O(nm \log^2(nm))$ for a $n \times m$ matrix and in $O(n\log(n))$ for a vector of length $n$. 
Assuming the ability to perform quantum queries to the entries of this data structure in superposition (i.e., assuming the availability of a QRAM that stores the entries of these trees), the authors show how to provide quantum access to a vector/matrix in time $\widetilde{O}(1)$.

In practice, in this input model, at the cost of some classical pre-processing, it is possible to encode vectors and matrices in quantum states using a small number of qubits in time $\widetilde{O}(1)$.

\subsection{\label{subsec:quantsubroutines}Relevant quantum subroutines}
We now introduce two quantum subroutines that are particularly important to our work. 
By slightly modifying these two results, we introduce two new corollaries that are more suitable to our needs.

\begin{lemma}[Inner product estimation \cite{kerenidis2019qmeans}]
\label{lem:innerprod}
Let there be quantum access to the matrices $V \in \mathbb{R}^{n \times m}$ and $C \in \mathbb{R}^{k \times m}$, through the unitaries
$U_v: \ket{i}\ket{0} \mapsto \ket{i}\ket{v_{i, \cdot}}$ and $U_c: \ket{j}\ket{0} \mapsto \ket{j}\ket{c_{j,\cdot}}$, that run in time $T$.
Then, for any $\delta > 0$ and $\epsilon>0$, there exists a quantum algorithm that  computes
$\ket{i}\ket{j}\ket{0}  \mapsto   \ket{i}\ket{j}\ket{\overline{\braket{v_{i,\cdot},c_{j,\cdot}}}}$,
such that
$| \overline{\braket{v_{i,\cdot},c_{j,\cdot}}}-\braket{v_{i,\cdot},c_{j,\cdot}} | \leq  \epsilon$,
with probability greater than $1-2\delta$
in time $\widetilde{O}\left(\frac{T\log(1/\delta)}{ \epsilon}\right)$.
\end{lemma}

In our quantum matching pursuit we will need to perform inner products between the columns of a matrix and a vector.
It is possible to use this result to prepare a state that stores the inner products between the columns of a matrix $A$ and a column vector $x$. 
Indeed, we only need quantum access to the matrix's transpose $A^T$ and to the vector $x$ via unitaries $U_A$ and $U_x$. 
If we have that, the two unitaries that prepare the states trivially become $U_v = U_{A^T}$ and $U_c = (\I \otimes U_x)$, and we can ignore the second register.
We also stress that having quantum access to a matrix is equivalent to having access to its transpose. 
Indeed, if we swap the first and the second register of Equation \ref{eq:matrice}, we have a quantum state that represents the transposed matrix.
Finally, using consistent phase estimation from \citet{ta2013inverting}, it is possible to modify the algorithm above so that the error across several runs is consistent. 

\begin{corollary}[Matrix-vector product estimation] \label{coro:inner}
Let there be quantum access to a matrix $A \in \mathbb{R}^{n \times m}$ and to a unit vector $x \in \R^m$, through the unitaries
$U_A: \ket{0}\ket{0} \mapsto \frac{1}{\norm{A}_F} \sum_i^n \ket{i}\ket{a_{i,\cdot}}$ and $U_x: \ket{0} \mapsto \frac{1}{\norm{x}_2}\sum_i^m x_i\ket{i}$, that run in time less than $T$.
Then, for any $\delta > 0$ and $\epsilon>0$, there exists a quantum algorithm that  computes
$\ket{0}\ket{0} \mapsto \frac{1}{\norm{A}_F} \sum_i^n \ket{i}\ket{\overline{\braket{a_{i,\cdot},x}}}$,
such that
$| \overline{\braket{a_{i,\cdot},x}}-\braket{a_{i,\cdot},x} | \leq  \epsilon$ consistently across multiple runs,
with probability greater than $1-2\delta$
in time $\widetilde{O}\left(\frac{T\log(m/\delta)}{ \epsilon}\right)$. 
\end{corollary}

To prove the bound on the success probability of the corollary above, we can exploit the union bound, also known as Boole's inequality
\begin{align}
    P(\cup_i^m p_f(i)) \leq \sum_i^m p_f(i). 
\end{align}
The union bound states that, given a set of likely events, the probability that any one of them happens is lower than the sum of their individual probabilities. 

The run-time overhead of Lemma \ref{lem:innerprod} to bound the failure probability of one inner product with $p_f(i) \leq 2\delta$ is $O(\log(1/\delta))$.
Bounding the failure probability of one product by $p_f(i) \leq 2\delta'$, the probability $P_f$ that one of the $m$ inner products fail can be bounded by
\begin{align}
    P_f \leq \sum_i^m p_f(i) \leq 2m\delta'.
\end{align}
Choosing $\delta' = \frac{\delta}{m}$, we have that the algorithm succeeds with probability $1-2\delta$ with a run-time overhead of $O\left(\log\left(\frac{m}{\delta}\right)\right)$.

Finally, if $x$ is not a unit vector, we can multiply the result by $\norm{x}_2$ to get an estimate within $\epsilon\norm{x}_2$ error.

The second algorithm that we introduce enables searching for the minimum value of an unsorted array quadratically faster than what we can do classically.
\begin{lemma}[Finding the minimum \cite{durr1996findingminimum}]
\label{lem:finding-min}
Let there be quantum access to a vector $u \in [0,1]^N$ via the operation $\ket j \ket{0} \to \ket j \ket{ u_j}$ in time $T$.
Then, we can find the minimum $u_{\min} = \min_{j\in[N]} u_j$ and its index $j_{\min} = \argmin_{j\in[N]} u_j$ with success probability $1-\delta$ in time $O\left(T \sqrt N \log \left (\frac{1}{\delta}\right) \right)$.
\end{lemma}

This algorithm is built around the famous Grover's search algorithm \citep{grover1996fast}.
Grover's algorithm takes advantage of an oracle to mark the elements of the superposition that satisfy the search conditions.
An oracle is a function $f: \R \mapsto \{0,1\}$. 
The Finding the minimum routine uses Grover's search many times, using oracles of the type
\begin{align}
    f_i(j) = \begin{cases}1 & \mbox{if } u_j < u_i\\
    0 & \mbox{otherwise}
    \end{cases}.
\end{align}
By using the same algorithm described in \citet{durr1996findingminimum} with oracles
\begin{align}\label{eq:maxabsoracle}
    f'_i(j) = \begin{cases}1 & \mbox{if } |u_j| > |u_i|\\
    0 & \mbox{otherwise}
    \end{cases},
\end{align}
it is possible to find the maximum absolute value of an array in the same running time. 

\begin{corollary}[Finding the maximum absolute value] \label{coro:finding}
Let there be quantum access to a vector $u \in [0,1]^N$ via the operation $\ket j \ket{0} \to \ket j \ket{ u_j}$ in time $T$.
Then, we can find the maximum absolute value $u_{\max} = \max_{j\in[N]} |u_j|$ and its index $j_{\min} = \argmin_{j\in[N]} u_j$ with success probability $1-\delta$ in time $O\left(T \sqrt N \log \left (\frac{1}{\delta}\right) \right)$.
\end{corollary}

%%%%%%%%%%%%%%%%%%%%%%%%%%%%%%%%%%%%%%%%%%%%%%%%%%%%%%%%%%%%%%%%%%%%%%%%%%%%%%%%%%%%%%%%%%%%%
%%%%%%%%%%%%%%%%%%%%%%%%%%%%%%%%%%%%%%%%%%%%%%%%%%%%%%%%%%%%%%%%%%%%%%%%%%%%%%%%%%%%%%%%%%%%%

\section{\label{sec:quantMP}Quantum matching pursuit}
Now that we have introduced the necessary quantum background, we show how to construct the quantum matching pursuit algorithm.
The proposed algorithm closely follows its classical counterpart, but it takes advantage of the quantum routines presented in Section \ref{subsec:quantsubroutines}. The main idea is to exploit the quantum regime to speed up the sweep stage.

\subsection{Data access}
First, we need quantum access to the dictionary $D$. 
We recall that, without loss of generality, we can consider the columns of $D$ to be $\ell_2$ normalized vectors, meaning that the Frobenius norm of $D$ is $\norm{D}_F = \sqrt{m}$.
Recall that, with a pre-processing time of $O(nm\log^2(nm))$, we can create quantum access to the following quantum state in time $\widetilde{O}(1)$
\begin{align}\label{eq:quantdictionary}
    \ket{D} = \frac{1}{\sqrt{m}} \sum_j^m \ket{j}\ket{d_{j}},
\end{align}
where $d_j$ is the $j^{th}$ column of $D$ and $\ket{d_j}$ is a state-vector.
This pre-processing cost needs to be paid only once and allows applying the quantum matching pursuit on any signal $s \in \R^n$ that is sparse over $D$.
For this reason, we will not include this cost in our run-time analysis.

\begin{figure}[t]
\includegraphics[scale=1]{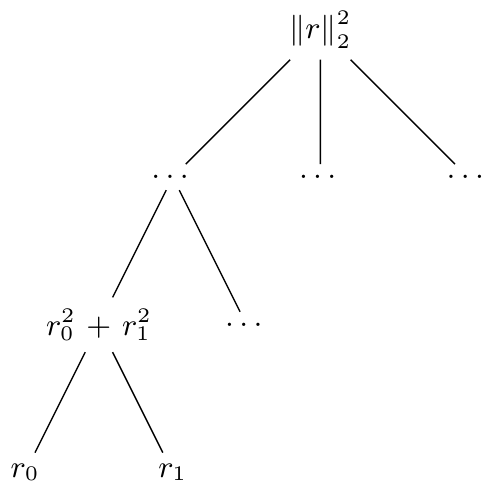}
\caption{The tree structure that enables efficient quantum access to the vector of the residuals. Each node stores the sum of squares of the leaves that descend from that node.}
\label{fig:tree}
\end{figure}

Similarly, we can create quantum access to the residual $r \in \R^n$ by using a tree data structure. 
For the sake of clarity, we report the data structure in Figure~\ref{fig:tree}.
This data structure can be created in time $O(n\log(n))$ for a vector with $n$ components and enables quantum access to the following quantum state in time $\widetilde{O}(1)$ \citep{kerenidis2017reccomender, grover2002discreteprobab}
\begin{align}\label{eq:quantresidue}
    \ket{r} = \frac{1}{\norm{r}_2} \sum_i^n r_i \ket{i}.
\end{align}

Preparing access to $\ket{r}$ and $\ket{D}$ means implementing the circuits described by the unitaries discussed in Section \ref{subsec:quantdatacces}.
Finally, both the signal $s \in \R^n$ and its sparse representation $x \in \R^m$ are represented as classical arrays, as they will not be encoded in quantum states. 
To further lower the memory complexity, it is possible to exploit the sparseness of $x$ and store it in a data structure which uses $O(\norm{x}_0)$ space (e.g., a hash table). 
This will not affect the overall time complexity as long as the data structure has a constant time insertion/update cost.

\subsection{Quantum algorithm}\label{subsec:quantalgo}
We start by initiating the residual structure with the signal components. 
Since we have quantum access to the residual and to the dictionary as in Equations \ref{eq:quantresidue} and \ref{eq:quantdictionary}, we can use the matrix-vector product estimation procedure from Corollary \ref{coro:inner} to produce the state
\begin{align}
    \ket{\varphi} = \frac{1}{\sqrt{m}}\sum_j^m\ket{j}\ket{\overline{z}_j}.
\end{align}
We recall that $z_j = \braket{d_j, r}$ and that the procedure computes $\overline{z}_j$ such that 
\begin{align}
    |\overline{z}_j - z_j|\leq \xi\norm{r}_2    
\end{align}
where $\xi \in \R^+$ is a parameter of arbitrary choice and $\norm{r}_2$ is the residual's $\ell_2$ norm at the current iteration.

Once we have a quantum register $\ket{\varphi}$ with a superposition of all the inner products, we can perform the Finding the maximum absolute value routine from Corollary \ref{coro:finding} to find $j^*$ and $z_{j^*}$.
With the best $j$ and $z_j$, we can proceed to update the solution $x \in \R^m$ and the residual $r \in R^n$. 

Just like in the classical algorithm, we repeat this procedure until the norm of the residual is lower than a threshold $\epsilon$ or the solution is such that $\norm{x}_0 > L$ for a threshold $L \in \mathbb{N}^+$.

Algorithm \ref{alg:qmp} concisely summarizes the quantum matching pursuit procedure.

\begin{figure}[t]
\begin{algorithm}[H]
    \caption{Quantum matching pursuit}
    \label{alg:qmp}
    \begin{algorithmic}[1]
        \State \label{alg:qmpinit}Initialize $r = s$, $x = 0^{\otimes m}$. 
        \While {not ($\norm{x}_0 > L$ or $\norm{r}_2 \leq \epsilon$)}
            \State \label{alg:qmpprepr}Prepare $\ket{r} = \frac{1}{\norm{r}_2} \sum_i^n r_i \ket{i}$
            \State \label{alg:qmpprepq}Prepare $\ket{D} = \frac{1}{\sqrt{m}} \sum_j^m \ket{j}\ket{d_j}$
            \State \label{alg:qmpinner}Use Inner product estimation to create  $\ket{\varphi} = \frac{1}{\sqrt{m}} \sum_j^m \ket{j}\ket{\overline{z}_{j}}$, where $z_j=\braket{d_j, r}$ and $|\overline{z}_j - z_j|\leq \xi\norm{r}_2$
            \State \label{alg:qmpfinding}Apply Finding the maximum absolute value to $\ket{\varphi}$ to obtain $j^*$ and $z_{j^*}$
            \State Update the solution $x_{j^*} = x_{j^*} + z_{j^*}$
            \State Update quantum access to the residual $r = r - z_{j^*} d_{j^*}$
        \EndWhile
    \State Output $x$.
    \end{algorithmic}
\end{algorithm} 
\end{figure}

It is important to remark that the error in the inner products can introduce convergence issues and could lead to worse solutions than the ones provided by the classical algorithm.
We have identified two alternative versions of the quantum matching pursuit algorithm:
\begin{enumerate}
    \item  \emph{Single-error}: The error only affects the computation of $j^*$. 
    We re-compute $z_{j^*}$ classically after the finding procedure outputs the atom's index;
    \item \emph{Double-error}: The error affects both the computation of $j^*$ and $z_{j^*}$, which are retrieved quantumly by the finding procedure.
\end{enumerate}
While the asymptotic complexity of the algorithm does not change between the two versions, the second one is slightly faster to compute.
In Section \ref{subsec:algcomp}, we provide experimental evidence that the two versions of the algorithm do not significantly affect the performance of quantum matching pursuit in practice, supporting the use of the second version.
We also provide numerical experiments to check whether the quantum algorithm can find representations of the same quality as its classical counterpart.

\subsection{Success probability}
Differently from its classical counterpart, the quantum algorithm that we propose has a probability of failing.
In this section, we discuss the success probability of our algorithm, showing that we can arbitrarily bound its failure probability with little running time overhead. 

The probability of failure is due to the computation of the inner products at step \ref{alg:qmpinner} and the search at step \ref{alg:qmpfinding}.
In particular, Corollaries \ref{coro:inner} and \ref{coro:finding} fail with a probability smaller than $2\delta_1'$ and $\delta_2'$, with a run-time overhead of $O(\log(\frac{m}{\delta_1'}))$ and $O(\log(\frac{1}{\delta_2'}))$ respectively.

Let us denote the probability of failure
of the $i^{th}$ loop iteration by $p_f(i)$ and consider $\delta_1'=\delta_2'=\delta'$.
The probability of success of the $i^{th}$ loop iteration is:
\begin{align}
    1-p_f(i) \geq (1-2\delta')(1-\delta') = 1 -3\delta' + 2\delta'^2 \geq 1-3\delta'
\end{align}
with a run-time overhead of $O(\log(\frac{m}{\delta'})\log(\frac{1}{\delta'})) = O(\log(\frac{m+1}{\delta'}))$.
So, the probability of failure of the $i^{th}$ loop iteration is bounded by $3\delta'$.

Given that the loop is executed $k$ times, by union bound, we can bound the total probability of failure as 
\begin{align}
    p(\cup_i^k p(i)) \leq \sum_i^k p(i) \leq 3k\delta'.
\end{align}
It follows that the success probability of our algorithm is $1 - 3k\delta'$ with a run-time overhead of $O(\log(\frac{m+1}{\delta'}))$.

By choosing $\delta' = \frac{\delta}{3k}$, we can affirm that our algorithm succeeds with probability greater than $1-\delta$ with a run-time overhead of 
$O(\log(\frac{3k(m+1)}{\delta}))\sim O(\log(\frac{3km}{\delta}))$.

Using the same proof technique, we could choose two different probabilities of failure $\delta_1'$ and $\delta_2'$ and further reduce the overhead by a constant factor. 
For instance, by choosing $\delta_1'=\frac{\delta}{2k}$ and $\delta_2'=\frac{\delta}{2km}$, we can bound the run-time overhead with $O(\frac{2km}{\delta})$.
Finally, note that even if the exact value of $k$ is not known beforehand, we can find suitable $\delta_1'$ and $\delta_2'$ by considering $k\sim O(L)$.

\subsection{Running time}
Finally, we provide a thorough analysis of the run-time of the new proposed algorithm, proving its computational complexity. 
As already said, we assume that the dictionary has been stored in an appropriate data structure, and we analyze the time to compute the sparse representation of a signal over that dictionary.

The first step of the algorithm consists in initializing the residual. 
This can be done by building the data structure in Figure~\ref{fig:tree}, and it requires time $O(n\log(n))$.
Since the residual and the dictionary are stored in adequate data structures, the preparation of states $\ket{D}$ and $\ket{r}$ at steps \ref{alg:qmpprepr} and \ref{alg:qmpprepq} is $\widetilde{O}(1)$, as discussed in Section \ref{subsec:quantdatacces}.

Given that the cost of preparing the two quantum states is $\widetilde{O}(1)$, from Corollary \ref{coro:inner}, we know that the cost of step \ref{alg:qmpinner} is $\widetilde{O}(\frac{1}{\xi})$ and that it encodes in the register values $\overline{z}_j$, with error $|\overline{z}_j - z_j| \leq \xi\norm{r}_2$.
Therefore, we can prepare the state 
\begin{align}
    \ket{\varphi} = \frac{1}{\sqrt{m}} \sum_j^m \ket{j}\ket{\overline{z}_{j}}
\end{align} 
in time $\widetilde{O}(\frac{1}{\xi})$.

We can ignore the run-time terms that depend on $\delta$ as they are related to the success probability studied in the previous section. 
We will include this overhead in the run-time at the end of this section.

Recall that the state above is the superposition of $m$ inner products, among which we need to identify the one with the greatest maximum absolute value.
Using Corollary \ref{coro:finding}, we can find the value $z_{j^*}$ and its index $j^*$ in time $O(\frac{\sqrt{m}}{\xi})$.

Once these values are computed, we need to update the solution and the residual. 
To update the solution, we need to add $\overline{z}_{j_*}$ to one component of the solution vector. 
We can do it in time $O(1)$.
On the other hand, updating the residual is more demanding, as we need to modify the tree data structure. 
Each element update costs $O(\log(n))$, as we need to update one leaf and all its parent nodes.
Since the update is $r = r + z_{j^*}d_{j^*}$, we need to update $\norm{d_{j^*}}_0$ elements of the residual. 
This step costs $O(\norm{d_{j^*}}_0\log(n))$.

Let us denote by $d_{j^*}^{(i)}$ the best atom at iteration $i$.
The cost of the $i^{th}$ iteration of the loop, results in
\begin{align}
    \widetilde{O}\left(\|d_{j^*}^{(i)}\|_0\log(n) + \frac{\sqrt{m}}{\xi}\right).
\end{align}
By assuming that we perform $k$ iterations of the loop, considering the initial cost of initiating the residual, and introducing the term that accounts for the success probability, the total complexity of our algorithm is
\begin{align}\label{complexity:exact}
    \widetilde{O}\left(n\log n + \sum_i^k \|d_{j*}^{(i)}\|_0 \log(n) + k\frac{\sqrt{m}}{\xi}\log\left(\frac{3km}{\delta}\right)\right).
\end{align}

If the atoms of our dictionary are not sparse, we can write the run-time in a more compact way by observing that $\forall j$, $\norm{d_{j}}_0 \leq n$
\begin{align}
    \widetilde{O}\left(kn\log n + k\frac{\sqrt{m}}{\xi}\log\left(\frac{3km}{\delta}\right)\right).
\end{align}

We can concisely summarise our result in the following theorem.
\begin{theorem}[Quantum matching pursuit]
Let there be quantum access to a dictionary $D \in \R^{n \times m}$ and to a vector $s \in \R^n$. Let $\xi, \delta \in \R_{>0}$ be precision parameters. 
There exists a quantum algorithm that simulates the matching pursuit algorithm, with error $\left|z_j^{(i)} - \overline{z}_j^{(i)}\right| \leq \xi\norm{r^{(i)}}_2$ on the inner products estimation at each $i^{th}$ iteration, in running time $\widetilde{O}\left(kn\log n + k\frac{\sqrt{m}}{\xi}\log\left(\frac{3km}{\delta}\right)\right)$, where $k$ is the total number of iterations. The algorithm succeeds with probability greater than $1 - \delta$.
\end{theorem}

%%%%%%%%%%%%%%%%%%%%%%%%%%%%%%%%%%%%%%%%%%%%%%%%%%%%%%%%%%%%%%%%%%%%%%%%%%%%%%%%%%%%%%%%%%%%%
%%%%%%%%%%%%%%%%%%%%%%%%%%%%%%%%%%%%%%%%%%%%%%%%%%%%%%%%%%%%%%%%%%%%%%%%%%%%%%%%%%%%%%%%%%%%%

\section{\label{sec:experiments}Numerical experiments}
The quantum algorithm that we propose executes the exact steps of its classical counterpart, but it introduces some random error along the computation, possibly affecting the quality of the representation. 
We present numerical simulations on synthetic data to better study the run-time advantage and the convergence properties of the quantum matching pursuit.
We study whether the \emph{Single-error} and \emph{Double-error} versions of the algorithm, introduced in Section \ref{subsec:quantalgo}, actually affect the algorithm's correctness.
Our numerical simulations are performed on a classical computer by introducing artificial errors $\xi \in [-0.01, 0.01]$ in the computation of the inner products.

\subsection{\label{subsec:algcomp}Single/double-error and representation quality}

We run experiments to test whether the two implementations of the quantum matching pursuit algorithm produce different results and whether the representation quality is different from the one obtained by the classical matching pursuit.

To do so, we proceed by creating $100$ batches, each of which composed of:
\begin{itemize}
    \item a random dictionary $D \in \R^{100\times512}$, with unit columns
    \item $100$ sparse vectors $x \in \R^m$, with $\|x\|_0=17$
    \item $100$ signals of the form $s = Dx + \epsilon$, where $\epsilon$ is a vector containing Gaussian truncated noise. %\footnote{}
\end{itemize}

The dictionary, the signals and the sparse vectors are provided by Scikit-Learn's~\citep{pedregosa2011scikit} \emph{make\_sparse\_coded\_signal} function. 
We add the $\epsilon$ noise artificially.

We simulate the two versions of the quantum matching pursuit algorithm and the classical one for each batch, to compare the sparse representations on the same data. 
In order to assess the convergence properties, we do not set any threshold $L$ that limits the sparsity of the solution.

We observe that all three versions represent the signals with $18$ components on average.
For both the \emph{Single-error} and the \emph{Double-error} version, we compute, for each batch, the average sparsity of the solution for that batch, divided by the one obtained using the classical version. The average value for this metric is respectively $1.0048$ and $1.0060$, indicating that the quantum and classical algorithms compute solutions with similar sparsity, and also that the two quantum versions do not differ much.

To better prove the latter point, we run a statistical test. First, we check whether the metric values obtained for each quantum algorithm version are normally distributed, using a Shapiro-Wilk normality test. 

The resulting $p$-values are respectively $0.37490$ for the \emph{Double-error} version and $0.07854$ for the \emph{Single-error} one. Since the \emph{Single-error} results are not normally distributed, and the experiments have been conducted on the same batches of data, we run a Wilcoxon signed-rank test to determine whether the performances are different. The test outcome is a $p$-value of $0.34790$, which does not allow us to reject the hypothesis that the two algorithms perform similarly.

\subsection{Run-time simulation}
While it is true that the classical complexity of the matching pursuit, in the general case, scales as $O(knm)$, and that the quantum version scales as $O(kn\log(n) + k\frac{\sqrt{m}}{\xi}\log(\frac{3km}{\delta})\text{polylog}(nm))$, it is legitimate to wonder whether the number of iterations $k$ is the same for both the classical and the quantum version.

To analyze the run-time of the algorithms, we compare the classical matching pursuit with the \emph{Double-error} quantum matching pursuit on the same dataset.

We generate datasets in the same way discussed in the previous section, with the difference that we study the run-time as the length $n$ of the signal increases.
For each batch, the number of atoms $m$ is set to twice the length of the signal, while solution is five times sparser than the original signal.

\begin{figure}[tbp]
\includegraphics[width=\linewidth]{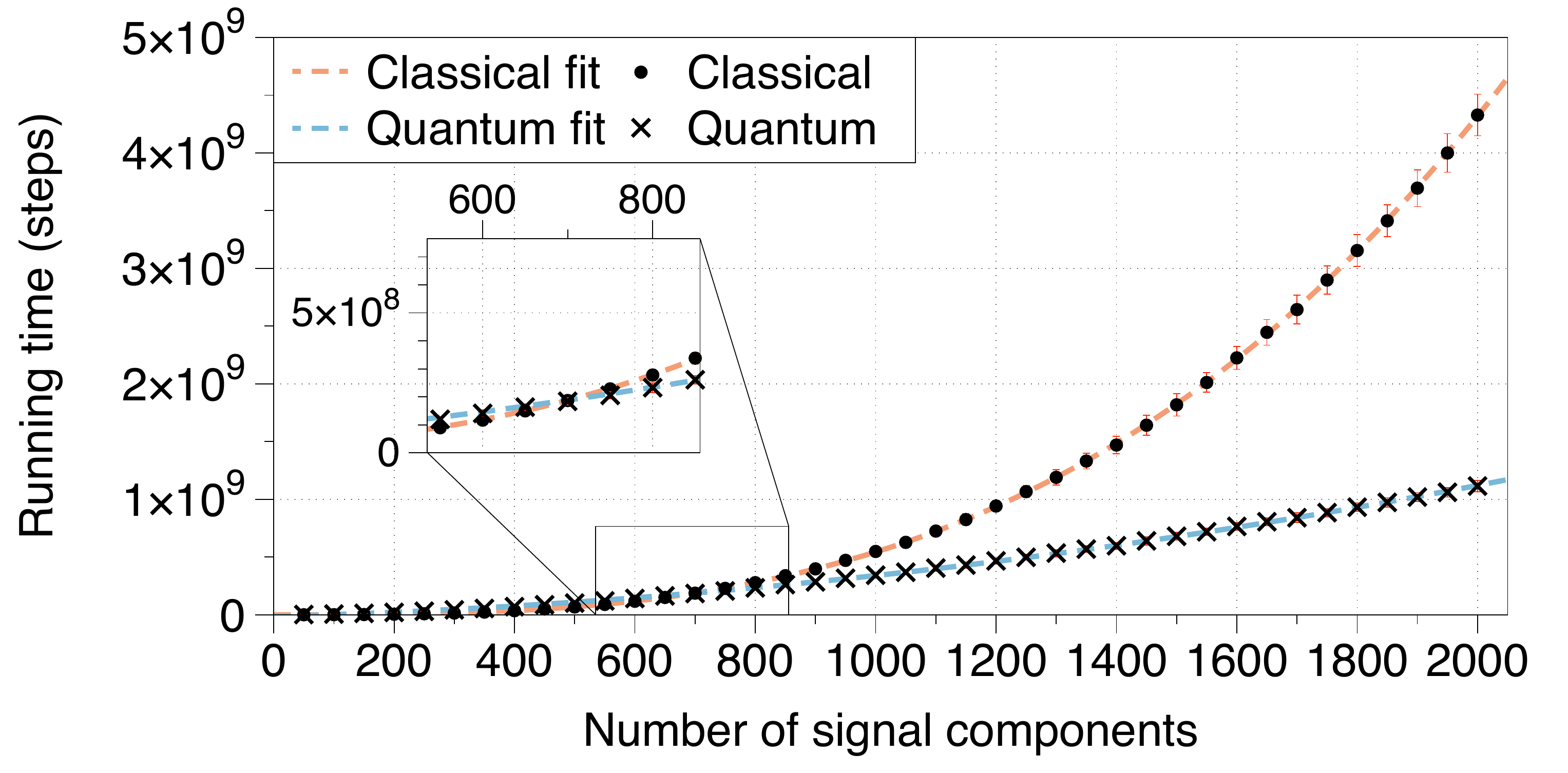}
\caption{\label{fig:qmpruntime} Running times for classical and quantum matching pursuit. Each point is the average number of operations required to find the sparse representation of a signal $s \in \R^n$. We report the standard deviation with red bars.}
\end{figure}

Experimentally, we see that there is not a significant difference in the number of iterations required to converge.
Indeed, the two algorithms produce solutions of the same sparsity.
Figure~\ref{fig:qmpruntime} illustrates the run-times of the classical and quantum algorithms, considering a probability of failure $\delta = 0.01$. 
Each point represents the average number of operations required to find the sparse representation of the signal, while the red bars show the standard deviation. 

We notice that the quantum algorithm is not advantageous until a certain number of signal components.
While the exact signal length for quantum advantage depends on the characteristics of the problem (e.g., number of atoms, expected sparsity), this experiment shows that the quantum matching algorithm provides a substantial speed-up over its classical counterpart on high-dimensional signals.

%%%%%%%%%%%%%%%%%%%%%%%%%%%%%%%%%%%%%%%%%%%%%%%%%%%%%%%%%%%%%%%%%%%%%%%%%%%%%%%%%%%%%%%%%%%%%
%%%%%%%%%%%%%%%%%%%%%%%%%%%%%%%%%%%%%%%%%%%%%%%%%%%%%%%%%%%%%%%%%%%%%%%%%%%%%%%%%%%%%%%%%%%%%

\section{\label{sec:conclusions}Conclusions}
In this work, we have presented a novel quantum algorithm to find sparse representations of signals in the QRAM input model.
The new algorithm is a quantum version of the classical matching pursuit algorithm.
We have leveraged, modified, and combined two quantum subroutines to speed up the sweep stage of the matching pursuit.
The result is a new routine with a polynomial advantage over its classical counterpart while retrieving solutions of the same quality.

This work shows that quantum computing can impact learning sparse representation, and opens the path to further research on other quantum pursuit algorithms.

\begin{acknowledgments}
The authors would like to thank Dr Alessandro Luongo for his valuable advice, and professors Ferruccio Resta and Donatella Sciuto for their support.
They are particularly grateful to Prof. Giacomo Boracchi for his inspiring lectures on sparse representations, and to Prof. Timothy J. Sluckin for his precious feedback on the first draft of this manuscript.
\end{acknowledgments}

\providecommand{\noopsort}[1]{}\providecommand{\singleletter}[1]{#1}%

\end{document}